\documentclass[english,prb,floats,superscriptaddress,usenames]{revtex4}
\usepackage[T1]{fontenc}
\usepackage[latin9]{inputenc}
\usepackage{amsmath}
\usepackage{amssymb}
\usepackage{graphicx}
\usepackage{esint}

\makeatletter

\newcommand{\be}{\begin{equation}} 
\newcommand{\ee}{\end{equation}}
\newcommand{\bea}{\begin{eqnarray}}   
\newcommand{\eea}{\end{eqnarray}}

\newcommand{\rr}{{\bf r}}

\newcommand{\eeta}{\boldsymbol{\eta}}
\newcommand{\txi}{\boldsymbol{\xi}^t}
\newcommand{\F}{\boldsymbol{F}}
\newcommand{\colxi}{\boldsymbol{v}}
\newcommand{\Fu}{{\cal F}}
\newcommand{\Ha}{{\cal H}}

\newcommand{\UU}{{\cal U}}

\usepackage{babel}

\makeatother

\usepackage{babel}
\begin{document}

\date{\today}

\title{Towards a statistical mechanical theory of active fluids}

\author{Umberto Marini Bettolo Marconi}
\address{ Scuola di Scienze e Tecnologie, 
Universit\`a di Camerino, Via Madonna delle Carceri, 62032, Camerino, INFN Perugia, Italy}
\author{Claudio Maggi}
\address{Dipartimento di Fisica, Universit\`a di Roma Sapienza, I-00185, Rome, Italy}

\begin{abstract}
We present a stochastic description  of a model of $N$ mutually interacting active particles in the presence of external fields  and characterize its steady state behavior  in the absence of currents. To reproduce the effects of the experimentally observed
 persistence of the trajectories of the active particles we consider  a Gaussian forcing having a non vanishing correlation time $\tau$,
 whose finiteness is a measure of the activity of the system.
With these ingredients we show that it is possible  to develop a statistical mechanical approach similar to the one employed in the study of equilibrium liquids and to obtain  the explicit form of the 
many-particle distribution function by means of the multidimensional unified colored noise approximation.
Such a distribution plays a role analogous to the Gibbs distribution in equilibrium statistical mechanics and provides a 
complete information about the  microscopic  state of the system.  From here we develop a method
to determine  the one and two-particle distribution functions in the spirit of
 the Born-Green-Yvon (BGY) equations of equilibrium statistical mechanics.  The resulting equations which
contain extra-correlations induced by the activity  allow to determine the stationary density profiles
in the presence of external fields, the pair correlations and 
the pressure of active fluids. In the low density regime we obtain the effective pair potential $\phi(r)$ acting between two isolated particles separated by a distance, $r$, showing the existence an effective attraction
 between them induced by activity.
 Based on these results,
 in the second half of the paper  we propose a mean field theory as an  approach simpler than the BGY hierarchy
 and use it to derive a van der Waals expression of the equation of state.
\end{abstract}

\maketitle

\section{Introduction}

The collective behavior of microscopic  living organisms  and  active  particles capable of transforming chemical energy
into motion  has recently attracted the attention of the soft-condensed matter community as they present striking analogies, but also
intriguing differences with respect to colloidal and molecular fluids \cite{romanczuk2012active, elgeti2014physics,marchetti2013hydrodynamics,ramaswamy2010mechanics}. Can we understand their rich phenomenology
by applying experimental, theoretical and numerical techniques which proved to be successful
in condensed matter physics? Recent investigations based on physical models akin
to those widely employed in statistical mechanics, thermodynamics and rheology
seem to give support to this hypothesis.

 Among the  most studied systems we mention  swimming bacteria,  colloidal  particles immersed in bacterial
 suspensions and self-propelled Janus particles.  Swimming bacteria  can be schematized as objects
moving with speed $v_0$ along straight paths of average duration $\tau$, the so called persistence time, after which a random reorientation (a tumble)  takes place. After many such events, i.e. on  a  time scale larger than $\tau$, the particle
displacement can be assimilated to a random walk and diffusive behavior emerges
with a diffusion constant, $D_a=v_0^2 \tau/2d$, where $d$ is the space dimensionality.  The resulting motion displays two peculiar features: a) the trajectories display an anomalously long persistence not observed in Brownian motion, 
i.e. their direction and the velocity remain constant for a time lapse much longer than those corresponding to colloidal particles,
and the similarity with ordinary diffusion appears only on a longer time-scale; b) there is a spontaneous tendency of the particles to aggregate into clusters notwithstanding
there is no evidence of direct attractive forces, while on the contrary short repulsive inter-particle forces are at work.
This is in a nutshell the idea of the successful run-and-tumble (RnT) model that 
    captures many aspects experimentally observed \cite{berg2004coli,schnitzer1993theory,tailleur2008statistical}. 
Experiments conducted employing  swimming bacterial suspensions   have shown that their diffusivity can be hundred times larger than the one arising from the 
thermal agitation \cite{valeriani2011colloids,di2010bacterial,schwarz2012phase}. To give some numbers the diffusivity of {\it Escherichia coli} bacteria  
is $D\approx 100\, \mu m^2/s$, whereas passive bacteria diffuse with $D\approx 0.3\, \mu m^2/s$. Active Janus particles instead have $D$ in a range   $4-25\, \mu m^2/s$,
finally colloidal particles immersed in a bacterial suspension display $D \approx 0.5-0.1\, \mu m^2/s$. In all cases, the contribution due to
the diffusion due to the thermal agitation of the molecules of the solvent is more than  ten times smaller than the one due to the activity
\cite{poon2013clarkia}.

The work of several groups has led to the formulation of theoretical models  and
 to the application of methods of  statistical mechanics \cite{cates2012diffusive,bialke2014negative,smallenburg2015swim,baskaran2010nonequilibrium,fily2012athermal,stenhammar2013continuum},
 whereas	Brady and Takatori  \cite{takatori2014towards} have put forward an approach based on  thermodynamics and microrheology and introduced the 
seminal idea  of swim temperature $T_s$ and swim pressure $T_s\rho$,  $\rho$ being  the particle number density.
In addition to the swim pressure, the total pressure contains other contributions stemming not only from the particle excluded volume
but also from the effective attraction experienced by the active particles.  Such an effective attraction is a peculiar
aspect of active matter:  motile particles, characterized by isotropic repulsive interactions due to the persistence of their motions become slower at  high density and tend to form clusters
and/or pile-up in  regions adjacent repulsive substrates.
The attractive forces may eventually lead to a scenario where  van der Waals loops appear 
as  soon as the off-equilibrium control parameter, the persistence time, is above a certain  critical value. 

The idea of the present paper is to use a microscopic  model capable of reproducing
 the basic features of the RnT model and of the active Brownian particles. In our model the microscopic state is specified only by the particles positions, while the effect of the angular dynamics  is encapsulated in a Gaussian colored noise having a finite relaxation time, reflecting the persistence of the directions. The correspondence between the models can be established by noting that, in the free particle case, in all models each velocity component has an exponential time-autocorrelation function characterized by a correlation time  $\tau$ ~\cite{koumakis2014directed,maggi2014generalized,maggi2015multidimensional,farage2015effective}.
Despite the differences in the driving stochastic forces, as shown in Ref.~\cite{maggi2015multidimensional}, Gaussian-colored noise driven particles behave strikingly similarly to RnT particles when in presence of steeply repulsive interactions. Moreover the Gaussian colored noise model is the only one for which an approximate stationary distribution for multiple particles is known~\cite{maggi2015multidimensional}. This motivates  us further in studying the Gaussian colored noise model in more detail.

Somehow surprisingly the present work shows that the approximate stationary configurational distribution of a system of  interacting particles undergoing over-damped motion under the action of Gaussian colored noise bears a strong resemblance with the equilibrium  Gibbs distribution \cite{tailleur2008statistical,cates2012diffusive}, with the due differences: 
a new temperature, named  swim temperature, takes  the role of the ordinary temperature, the probability of configurations depends    on a
complicated function of $\UU$, the  potential energy.
Such a result is independent on the fine details of the interactions, but certainly depends on the persistence of the random fluctuations.
 The most direct consequence of the form of this steady state distribution is the appearance of an effective attraction between two or more active
 particles  in the absence of any attractive direct interaction~\cite{angelani2011active}.
 The reduction of the mobility of the particles due to the presence of other particles in their vicinity may eventually lead to a phase separation between a high and a rarefied density phase, a phenomenon named motility induced phase separation (MIPS)~\cite{tailleur2008statistical,cates2014motility}.

This paper is organized as follows:  in Sec. \ref{Model} we present  the coarse grained stochastic model 
describing an assembly of active particles, consisting of a set of coupled Langevin equations for the coordinates of the particles
subject to colored Gaussian noise. We switch from the Langevin description to  the corresponding Fokker-Planck equation 
for the joint probability distribution of $N$ particles and within the multidimensional unified colored noise approximation 
	(MUCNA) \cite{hanggi1995colored,maggi2015multidimensional},   in the stationary case and under the condition of vanishing currents we obtain its exact form.
 The obtained distribution  implies detailed balance  and potential conditions as discussed in appendix C. In the case of non vanishing currents it is straightforward to write such a distribution for a single particle in one dimension in the presence of colored noise  \cite{lindner1999inertia}, whereas in higher dimensions and in interacting systems sucha a generalization cannot be obtained by the method
of appendix A.
Using the  $N$-particle distribution in Sec. \ref{BGY} we derive the first two members of a hierarchy of equations for the marginalized probability
distributions of one and two particles that play the role of the Born-Green-Yvon (BGY)  equations for active systems.
We apply the first of these equations to the study of a density profile in the presence of an external potential, 
extend the treatment to the case of interacting active particles.
In the low density limit we are able to write the exact form of such the pair correlation and define the effective pair potential
 between two isolated particles.
In section \ref{MFT} we employ a variational approach,  complementary  to that based on the BGY equations
which are exact but highly unpractical,  and  taking advantage of the explicit form of the MUCNA equation we construct
a "free energy" functional whose minimum corresponds to the exact solution.
 The  mean-field theory for interacting systems  follows by searching the solution among the probability distributions which are a product
 of single particle probability distributions.
  The functional is finally used in section \ref{vdW} to lay out a method capable to interpret the
phase behavior of the model in terms of the relevant control parameters starting from a microscopic description.
Finally, we present our conclusions in \ref{conclusions}.
We include with four  appendices: in appendix   \ref{appendixA} containing the calculation details leading to the MUCNA,
 in appendix  \ref{Htheorem} we discuss the approach to the solution and establish an H-theorem, in appendix \ref{detailedbalance} we verify the detailed balance condition,
whereas  in appendix \ref{appendixC} we discuss a technical aspect which allows  to evaluate a key ingredient
of our approach, the determinant of the friction matrix. 


\section{The model and its stationary many-particle distribution function}
\label{Model}

In order to describe the properties of a suspension of active particles
we consider a three dimensional container of volume $V$ where an assembly of  $N$ interacting active spherical particles at positions $\rr_i$, with $i$ ranging from 1 to $N$,
 subject to external fields undergoes over-damped motion driven by random fluctuating forces of different
 origin and nature.  In fact, each particle besides experiencing a white noise force, due to the thermal agitation of the solvent and characterized by a diffusivity
 $D_t$ , is acted upon by a drag force proportional to its velocity, $\dot \rr_i$,  and in addition is  subject to a colored Gaussian noise,
 $\colxi_i$, of zero mean, characteristic time $\tau$ and diffusivity $D_a$. 
 This second type of noise is intended to mimic on time scales larger than $\tau$  the behavior of self-propelled particles
 whose propulsion force randomizes with characteristic  time $\tau$. 
 Such a model involving only  the positional degrees of freedom of the particles has the advantage of allowing for analytical progress and for this reason can be used instead of more microscopic models where the rotational dynamics is fully accounted for.
We consider the following set of equations of motion for the positions of the active Brownian spheres which has been treated in the literature by some authors \cite{farage2015effective,elgeti2014physics,maggi2015multidimensional} :
\be
\label{effective_langevin}
\dot{\rr}_i(t) =\frac{1}{ \gamma}\F_i(\rr_1,\dots,\rr_N) +D_t^{1/2} \txi_i(t) + \colxi_i(t) 
\ee
are coupled to changes of the velocities $\colxi$ described by:
\be
\dot\colxi_i(t) =- \frac{1}{\tau}\colxi_i(t) + \frac{ D_a^{1/2}}{ \tau} \eeta_i(t)  
\label{exp_correlation}
\ee 
The force $\F_i= -\nabla_i \UU$ acting on the $i$-th particle is  conservative and associated to
the potential $\UU(\rr_1,\dots,\rr_N)$, $\gamma$ is the drag coefficient, whereas
the stochastic vectors $\txi_i(t)$ and $\eeta_i(t)$ are Gaussian and Markovian processes distributed with zero mean
and moments $\langle\txi_i(t)\txi_j(t')\rangle=2 \delta_{ij}\delta(t-t')$  and $
\langle \eeta_i(t) \eeta_j(t')\rangle=2  \delta_{ij}  \delta(t-t')
$.  
While $D_t$ represents a translational diffusion coefficient, the coefficient $D_a$ due to the activity is related to the correlation of the 
Ornstein-Uhlenbeck process  $\colxi_i(t)$ via 
$$\langle\colxi_i(t)\colxi_j(t')\rangle=2 \frac{D_a}{\tau} \delta_{ij}\exp( -|t-t'|/\tau) .$$ 
In spite of the fact that the magnitude of the velocity has  not the fixed value $v_0$ as in the RnT, but fluctuates,
a correspondence can be established between
 the present model  and the RnT by requiring that the mean  value  $\langle v_i^2\rangle=v_0^2$, so that
$D_a=v_0^2 \tau /2d $.

In the following we shall adopt a shorthand notation in which the set of vectors $\rr_i$ is represented by an array $\{x_i\}$ of dimension $M=dN$ and the remaining terms are replaced by non bold letters.
We assume that the force $F_i$ may be due to action of external agents and mutual interactions between the particles.
In order to proceed we consider the multidimensional version of the unified colored noise approximation (UCNA)\cite{hanggi1995colored} which consists in eliminating adiabatically the fast degrees of freedom of the problem and, as shown in appendix \ref{appendixA}, arrive at the following equation for the 
 the particles coordinates:
\be
\dot x_i \simeq \sum_k \Gamma^{-1}_{ik}  \Bigl[ \frac{1}{\gamma}  F_k+
D_a^{1/2}  \eta_k(t) + D_t^{1/2}     \xi^t_k  \Bigl]
\label{stochasticxi}
\ee
where we introduced the non dimensional friction  matrix $\Gamma_{ik}$
\be
\Gamma_{ik}=\delta_{ik}+\frac{\tau}{\gamma} \frac{\partial^2 \UU }{\partial x_i \partial x_k}.
\label{gammamatrix}
\ee
Such a formula says that the effective dynamics of each particle depend on its distance relative to the other particles and on its
absolute position
if an external field is present,
not only through the direct coupling $F_k$, but also through the motility matrix,
 $\frac{1}{\gamma} \Gamma^{-1}_{ik}$ which is the sum of a constant contribution due to the
background fluid plus a space dependent term due to the interparticle forces mediated by the colored bath.

 Notice that such a structure of the friction matrix introduces velocity correlations 
among  different velocity components of a given particle or between the velocities of different particles (this aspect will be discussed in detail in a
forthcoming publication).
For this reason, the present approach is not a mapping onto a passive equilibrium system, but rather 
represents a  mapping onto a system with a self generated inhomogeneous friction. This leads to strong deviations from equilibrium
such as the explicit dependence of the stationary state on the transport coefficient.

In order to obtain meaningful results one must require that all eigenvalues of $\Gamma$ are non negative.
We cannot prove such a condition in general, however, it seems a reasonable assumption for repulsive pair potentials.
On the contrary, it is easy to find examples where appropriately chosen external potentials determine negative eigenvalues and limit the validity of our formula.
	The structure of  eq. \eqref{stochasticxi} together with  \eqref{gammamatrix} is interesting because it tells that the damping
	experienced by  a particle is due to a standard drag force $-\gamma \dot x$ plus a contribution stemming from 
	interactions. Thus  the effective friction increases with density leading to lower mobility and to 
	a tendency to cluster. This mechanism can eventually lead to the (MIPS)~\cite{tailleur2008statistical,cates2014motility} and is an intrinsically non-equilibrium effect.

Let us write the Fokker-Planck equation (FPE)  for the $N$-particle distribution function $f_N$ associated  with the stochastic differential equation eq. \eqref{stochasticxi}
under  the Stratonovitch convention \cite{gardiner1985stochastic}
\bea
\frac{ \partial f_N(x_1,\dots,x_N;t)}{\partial t}=-\sum_{l}  \frac{\partial }{\partial x_l} J_l (x_1,\dots,x_N;t) 
\label{FPE}
\eea
where  the $l$-th  component  of the  probability current  is:
\bea
J_l\equiv\sum_k \Gamma^{-1}_{lk} \Bigl(\frac{1}{\gamma } F_k  f_N 
- (D_a+D_t) \sum_{j}      \frac{\partial }{\partial x_j}
 [ \Gamma^{-1}_{jk}   f_N ] \Bigr)
 \label{Ncurrent}
 \eea
 Using the method illustrated in Appendix \ref{appendixA} and requiring the vanishing of all components of the probability current vector $J_l$,  without further approximations we obtain the following set of conditions  for the existence of the steady 
state  $N$-particle distribution function $P_N$:
 \bea
 -(D_a+D_t)\gamma \Bigl(\frac{\partial P_N}{\partial x_n}-P_N   \frac{\partial}{\partial x_n} \ln \det \Gamma \Bigr)
     -P_N\sum_k  \Bigl( \delta_{nk}+\frac{\tau}{\gamma} \frac{\partial^2 \UU }{\partial x_n \partial x_k}  \Bigl)  
\frac{\partial \UU(x_1,\dots,x_N)}{\partial x_k} =0 .
\label{pformul2abis}
\eea


We  now define the effective configurational energy of the system $\Ha(x_1,\dots,x_N)$ related  to the bare potential energy
$\UU(x_1,\dots,x_N)$, by:

\be
\Ha(x_1,\dots,x_N) = \UU(x_1,\dots,x_N)  +\frac{\tau}{2 \gamma}
    \sum_k^N \Bigr(\frac{\partial \UU(x_1,\dots,x_N)}{\partial x_k} \Bigl)^2 
    - (D_a+D_t)\gamma  \ln  |\det   \Gamma_{ik}  |
\ee

It is easy to verify that  the following $N$-particles configurational  probability distribution, obtained for the first time
in ref. \cite{maggi2015multidimensional},
  \be
   P_N(x_1,\dots,x_N)  = \frac{1}{Z_N} \, \exp \Bigl(- \frac{\Ha(x_1,\dots,x_N)}{(D_a+D_t) \gamma} \Bigr)
   \label{probabilitydistr}
    \ee
    is a solution of eq. \eqref{pformul2abis}, where
      $Z_N$ is a normalization constant.
      the analogue of the canonical partition function and enforcing the condition
 $ \int\dots \int P_N(x_1,\dots,x_N) \, dx_1 \dots dx_N=1 $.
 Notice that the mobility $\Gamma^{-1}_{ik}$ enters the stationary distribution, at variance with equilibrium 
 systems where the kinetic coefficients never influence the form of the probability distribution function. 
   In the limit $\tau\to 0$, $Z_N$ reduces to  the equilibrium configurational partition function for a system characterized by energy $\UU$.
      Formula   \eqref{probabilitydistr} is a generalization to spaces  of arbitrary dimensions of a stationary distribution 
	  obtained by Hanggi and Jung for a single degree of freedom in one dimension \cite{hanggi1995colored}. It displays an exponential dependence on a 
  function constructed with the potential $\UU$ and its derivatives. These derivatives must be non singular and also
  satisfy the condition that the determinant is non negative.
  In the case of a single particle the result can also be derived by a multiple-time scale method
	  often employed to reduce the phase-space Kramers equation to the configurational space Smoluchowski equation \cite{marconi2006nonequilibrium}.
    Notice, that the explicit form of the distribution allows to identify an effective temperature of the system
    with the quantity 
    \be
    T_s=(D_t+D_a)\gamma \, ,
    \ee
    which shows the typical Einstein fluctuation-dissipation relation (with $k_B=1$)
    between temperature, diffusivity and and drag coefficient.
    Let us remark that
    the validity of  formula   \eqref{probabilitydistr}
is limited to the regime $D_a>D_t$, which  also corresponds to the actual 
 experiments with bacterial baths,  whereas the limit $D_a\to 0$ with $\tau$  and $D_t$ finite, neither corresponds to the real situation 
but also leads to meaningless theoretical predictions since it violates the hypothesis under which the MUCNA is obtained.
   To conclude this section, 
  we identify the temperature with $T_s$ with swim temperature   \cite{takatori2014swim}
  and introduce as a measure of the distance from thermodynamic equilibrium the P\'eclet number, $Pe=\sqrt{D_a\tau}/\sigma$, which is the ratio between  the mean square diffusive displacement due to the active bath in a time interval $\tau$ and
  the typical size of the active particles, say  $\sigma$.
  \section{BGY hierarchy and fluid structure}
 \label{BGY}
 As it stands formula \eqref{probabilitydistr} is exact, but contains too many details to be 
 really useful, however, borrowing methods of equilibrium statistical mechanics we can  trace out degrees of freedom and arrive
at formulas involving the distribution functions  of only  few particles. To this purpose we shall use
 the stationarity condition   \eqref{pformul2abis} to derive a set of equations
 equivalent to the BGY hierarchy for equilibrium correlations. The hierarchy
  becomes of practical utility if utilized in conjunction with a suitable 
 truncation scheme in order to eliminate the dependence from  the higher order correlations.
 Let us  turn to standard vector notation where the indices $\alpha$ and $\beta$ running from $1$ to $d$  identify the Cartesian components and latin subscripts the particles.
 The total potential is assumed to be the sum of the mutual pairwise  interactions $w(\rr-\rr')$ between the particles 
 and of the potential exerted by the external field $u(\rr)$:
 $\UU (\rr_1,\dots,\rr_N)=\sum_{i>j}^N w(\rr_i,\rr_j) +\sum_i^N u(\rr_i) .$
The hierarchy follows from writing eq. \eqref{pformul2abis}  in the equivalent form:
 \bea
  -T_s \sum_\beta \sum_n \frac{\partial }{\partial r_{\beta n}}[  \Gamma^{-1}_{\alpha l,\beta n}(\rr_1,\dots,\rr_N) P_N(\rr_1,\dots,\rr_N)]=    
   P_N(\rr_1,\dots,\rr_N) \Bigl(  \frac{\partial u(\rr_{\alpha l})}{\partial r_{\alpha l}} +  \sum_{k\neq l}   \frac{ \partial w(\rr_l-\rr_k) }{ \partial r_{\alpha l}}   
\Bigl)
\eea  
We proceed to marginalize the $N$ dimensional distribution function $P_N$  introducing the reduced probability distribution functions of order $n$ as
 $
P_N^{(n)}(\rr_1,x_2,\dots,\rr_n)\equiv \int d\rr_{n+1} \dots d \rr_N P_N(\rr_1,\rr_2, \dots,\rr_N)
$.
By integrating eq. \eqref{pformul2abis}
over  $(N-2)$ coordinates we obtain an equation for $P_N^{(2)}(\rr_1,\rr_2)$ in terms of higher order marginal distributions and choosing $n=2$ we find:
\bea
&&
  -T_s \int \int d\rr_3\dots d\rr_N \sum_\beta \sum_n \frac{\partial }{\partial r_{\beta n}}[  \Gamma^{-1}_{\alpha 1,\beta n}(\rr_1,\dots,\rr_N) P_N(\rr_1,\dots,\rr_N)]=    
   P^{(2)}_N(\rr_1,\rr_2) \Bigl(  \frac{\partial u(\rr_{\alpha 1})}{\partial r_{\alpha 1}} +  \frac{ \partial w(\rr_1-\rr_2) }{ \partial r_{\alpha 1}}  \Bigr)
   \nonumber\\
   && 
   + \sum_{k>2} \int   d\rr_k   P^{(3)}_N(\rr_1,\rr_2,\rr_k)   \frac{ \partial w(\rr_1-\rr_k) }{ \partial r_{\alpha 1}}   
\label{p2distr}
\eea  
Now, we notice that in the case of a large number of particles and in the limit of small $\tau/\gamma$ the matrix $\Gamma^{-1}_{\alpha 1,\beta n}$ is nearly diagonal and can be approximated by 
$$
\Gamma^{-1}_{\alpha l,\beta n}\approx \Bigl( \delta_{\alpha\beta}-\frac{\tau}{\gamma} u_{\alpha\beta}(\rr_l)-
\frac{\tau}{\gamma}  \sum_{k\neq l}  w_{\alpha\beta}(\rr_l-\rr_k) ) \Bigl)\delta_{ln},
$$ 
where $u_{\alpha\beta}\equiv \frac{\partial^2 u(\rr)}{\partial r_\alpha \partial r_\beta}$ and
$w_{\alpha\beta}\equiv \frac{\partial^2 w(\rr)}{\partial r_\alpha \partial r_\beta}$.
 Substituting this approximation in eq. \eqref{p2distr} we find:
\bea
&&
 T_s  \sum_\beta \frac{\partial }{\partial r_{\beta 1}}\Bigl[P^{(2)}_N(\rr_1,\rr_2)  \delta_{\alpha\beta} -    \frac{\tau}{\gamma}    \Bigl(P^{(2)}_N(\rr_1,\rr_2)
  u_{\alpha\beta}(\rr_1) + P^{(2)}_N(\rr_1,\rr_2)w_{\alpha\beta}(\rr_1-\rr_2)   + \int \sum_k d\rr_k  P^{(3)}_N(\rr_1,\rr_2,\rr_k) w_{\alpha\beta}(\rr_1-\rr_k) \Bigl)\Bigl]
  \nonumber\\
   && 
    =  -P^{(2)}_N(\rr_1,\rr_2) \Bigl(  \frac{\partial u(\rr_{\alpha 1})}{\partial r_{\alpha 1}} +  \frac{ \partial w(\rr_1-\rr_2) }{ \partial r_{\alpha 1}}  \Bigr)
    - \sum_{k>2} \int   d\rr_k   P^{(3)}_N(\rr_1,\rr_2,\rr_k)   \frac{ \partial w(\rr_1-\rr_k) }{ \partial r_{\alpha 1}}   ,
\label{p2bgy}
\eea  
which represents the BGY equation for the pair probability distribution $P_2$.
By integrating also over the coordinate $2$ and
switching to the $n$-th order density distributions:
$
\rho^{(n)}(\rr_1,\rr_2,\dots,\rr_n)= \frac{N!}{(N-n)!}  P_N^{(n)}(\rr_1,\rr_2, \dots,\rr_n) $
we obtain the first BGY-like equation:
\bea
&&
 T_s \sum_\beta \frac{\partial }{\partial r_{\beta 1}}\Bigl[ \delta_{\alpha\beta} \rho^{(1)}(\rr_1) -    \frac{\tau}{\gamma}    \rho^{(1)}(\rr_1)
  u_{\alpha\beta}(\rr_1) -\frac{\tau}{\gamma}  \int d\rr_2   \rho^{(2)}(\rr_1,\rr_2) \ w_{\alpha\beta}(\rr_1-\rr_2) \Bigl]
  \nonumber\\
   && 
    =  -\rho^{(1)}(\rr_1)   \frac{\partial u(\rr_{1})}{\partial r_{ \alpha 1}}
    - \int   d\rr_2   \rho^{(2)}(\rr_1,\rr_2)   \frac{ \partial w(\rr_1-\rr_2) }{ \partial r_{\alpha 1}}   ,
\label{rho1bgy}
\eea  
that in the limit of $\tau\to 0$  is just the  BGY equation for the single-particle distribution function.
By performing an analogous substitution in \eqref{p2bgy} we obtain the BGY equation for the pair correlation function
including  the corrections of order $\tau/\gamma$ stemming from the activity. 

The r.h.s.  of equation \eqref{rho1bgy}  contains the coupling to the external field and the so-called direct interaction among the particles,
whereas the l.h.s. besides the ideal gas term contains a term proportional to the activity parameter that we name 
indirect interaction term, following the nomenclature introduced by 
Solon et al. \cite{solon2014pressure,solon2014pressure2}, although our expression does not coincide with theirs  
because the present model  does not depend on angular
degrees of freedom.
Notice that here
 the indirect interaction term  stems from the expansion of the determinant to first order in the parameter $\tau/\gamma$.
 On the other hand, if one employed an higher order expansion in this parameter
  terms involving terms up to  $N$-body correlations would appear. 

\subsection{Non interacting active particles under in-homogeneous conditions} 

In the case of  vanishing "inter-molecular"  forces  equation \eqref{rho1bgy} gives access to the single particle distribution 
of Brownian active particles near a wall. Let us assume $w(r)=0$ and $u(\rr)=u(x)$ to represent a 
generic (flat) wall confining
potential, twice differentiable and with the properties that $\lim_{x \to -\infty} u(x)=\infty$ and  $\lim_{x \to \infty} u(x)=0$ .
Since the particles are non interacting it is not necessary to expand the friction matrix in powers of activity parameter as in eq. \eqref{rho1bgy}
and one can use its exact expression:
$$\Gamma^{-1}_{x l,x n}=\frac{1}{ 1+\frac{\tau}{\gamma} u_{xx}(x)}\delta_{ln}. $$
Alternatively, one can regard such a formula as a resummation of the  generalized binomial series $(1-y+y^2\dots)$,
whose first term, $y=\frac{\tau}{\gamma} u_{xx}(x)$  appears in the l.h.s. of \eqref{rho1bgy}.
We write:
  \be
T_s    \frac{d}{d x} \Bigl( \frac{  \rho^{(1)}(x)}{  1+ \frac{\tau}{\gamma}
   u_{xx}(x)   }\Bigr)=
-  u_{x}(x) \rho^{(1)}(x)
   \label{YBG1t} 
 \ee
and find the profile:
\be
\rho^{(1)}(x)=\rho_0 \exp \left(- \frac{u(x) + \frac{\tau}{2\gamma}
 ( u_x (x))^2  }{T_s} \right)  \Bigl[1+ \frac{\tau}{\gamma}
   u_{xx}(x)   \Bigl] \, .
 \label{walldistribution}
 \ee
  Using the mechanical definition of pressure \cite{solon2014pressure,winkler2015virial},
this simple example provides an exact expression for the 
pressure exerted by  non interacting  active particles:  in fact, by integrating both sides of \eqref{YBG1t}  with respect to $x$ from $-\infty$ to $\infty$ and recalling that
  the right hand side is just the force per unit area exerted by the wall, located at $x \approx 0$, on the fluid we obtain the pressure  as:
  \be
  p_s=T_s  \rho(x\to {\infty})
  \label{pressureideal}
  \ee
  having assumed that the negative region is not accessible to the particles. Notice that $p_s=\rho T_s$  does not depend on the particular form of the wall potential \cite{solon2014pressure}.
Such a pressure is the so called swim pressure that is the sum of the active  and passive ideal pressures.
 As remarked by Brady $p_s$ may depend on the size of the particles only through the hydrodynamic drag factor $\gamma$ and in the present case contains a thermal contribution associated with $D_t$ and an athermal part associated with the active dynamics \cite{takatori2014swim}.

Repulsive barriers with positive curvature ($u_{xx}>0$) induce  a local accumulation of particles 
 and lower  their mobility with respect to their bulk value. One can speculate that a similar phenomenon occurs spontaneously
in an interacting system   where denser and less motile clusters of active particles "attract" fast moving particles from
rarefied regions:  the flux would be sustained  by the difference between the pressures of the two regions. We shall
consider the role of interactions among the particles in the section below.

\begin{figure}[h]
\centering
  \includegraphics[height=10cm]{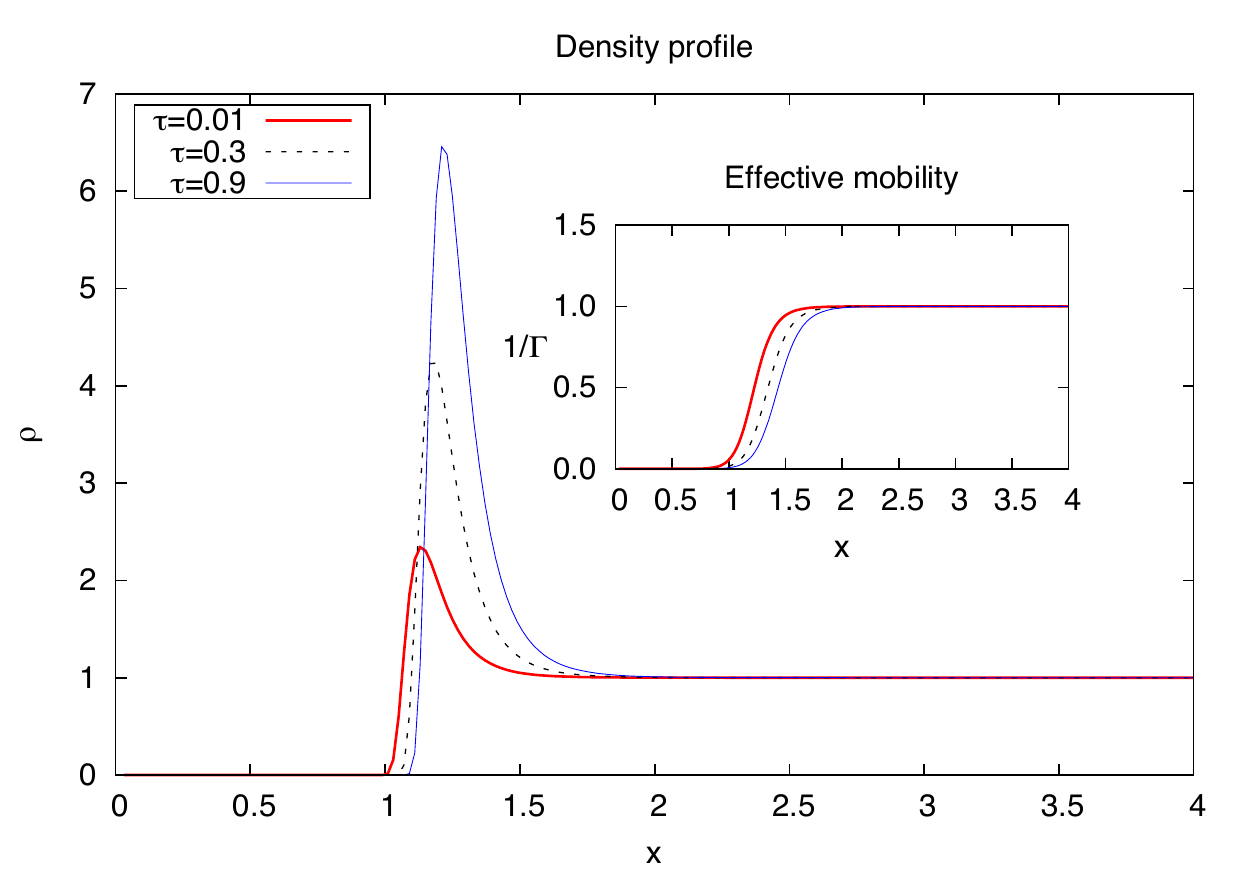}
  \caption{Three density profiles in the presence of a repulsive soft wall
  of the form $u(x) =u_0(\frac{\sigma}{x})^{12}$  corresponding  to three  different values of the 
  persistence time $\tau$ as indicated in the box. In the inset we report
  the effective mobility profiles for the same values of $\tau$. }
  \label{fgr:onedimension}
\end{figure}

Interestingly, we can  rewrite   \eqref{YBG1t}   as the local balance equation between a local pressure  term $\tilde p_s(x)= \tilde T_s(x)\, \rho^{(1)}(x) $
and a force term 
with  the local swim temperature $\tilde T_s(x)$  defined as
\be
 \tilde T_s(x)= \frac{T_s}{   1+\frac{\tau}{\gamma}  u_{xx}(x)  } .
 \label{tslocal}
\ee
In Fig. \ref{fgr:onedimension} we display both the density and temperature profiles next to a repulsive wall for three  different values of the
persistence time.
Formula \eqref{YBG1t} applies to rather general potentials under the condition that $u_{xx}(x)>-\gamma/\tau$. 

It is of interest to apply   \eqref{YBG1t}  to the special case  of sedimentation 
of active colloidal suspensions, where $u(x)=mgx$. One immediately finds the barometric law 
$\rho(x)=\rho_0\exp(- mg x/T_s)$, predicted by Tailleur and Cates in the small sedimentation rate limit \cite{tailleur2009sedimentation} and confirmed experimentally by Palacci et al. \cite{palacci2010sedimentation} who performed
the so-called  Perrin sedimentation experiment using active particles.
This scenario was also confirmed for swimming bacteria under centrifugation~\cite{maggi2013motility}. Notice that in this case the 
linear potential only appears in the exponent as if the system were at equilibrium, but with an effective
temperature $T_s$ higher then the ambient temperature $T_0=D_t\gamma$ according to the
formula $T_s/T_0=1+D_a/D_t$.
Formula \eqref{YBG1t}  is more general and encapsulates the idea that repulsive interactions render the diffusion
of the particles less efficient.
  
\subsection{Interacting active particles and non ideal pressure via BGY equation} 
To extend  eq.\eqref{YBG1t} to the case of interacting active particles we apply
an external potential varying only in the direction $x$ and  factorize the multiparticle distributions as $\rho^{(n)}(\rr_1,\dots,\rr_n)\approx  \rho^{(1)}(\rr_1) \dots \rho^{(1)}(\rr_n)$ invoking 
 a mean field-like  approximation and  recast  \eqref{rho1bgy} as:
 \bea
&&-T_s \frac{d}{dx_1} \left(\frac{  \rho^{(1)}(x_1)   }{ 1+ \frac{\tau}{\gamma}
 u_{x1,x1}(x_1)  
+  \frac{\tau}{\gamma} \int d\rr_2     
 \rho^{(1)}(\rr_2)   w_{x1,x1}(\rr_1,\rr_2))   }\right)
\approx
\nonumber\\
&& \rho^{(1)}(x_1) \frac{d}{d x_1} u(x_1)     +  \rho^{(1)}(x_1) \int d\rr_2 \rho^{(1)}(x_2)  \frac{\partial}{\partial x_1}   w(\rr_1,\rr_2))  
  \label{YBGsette}
\eea
where we have dropped off-diagonal terms $w_{\alpha\beta}$ in the denominator by the symmetry of the planar problem.

Formula  \eqref{YBGsette}   is valid up to linear order in  $\tau/\gamma$ and
in the l.h.s. contains the gradient of the density multiplied by a space dependent mobility, represented by the denominator. 
It generalizes to the  interacting case the denominator featuring in formula \eqref{YBG1t} by the presence of  inter-particle contributions.
When $w_{xx}(\rr)$  is positive it gives rise
to a  slowing down of particle motion due to the surrounding particles 
 in interacting active systems and determines a density dependence of the 
mobility.
In the r.h.s. the first two terms represent the external force term and the contribution to the pressure gradient due to direct "molecular" interactions, respectively.
Notice again that the denominator represents the combined action of the
"molecular" forces  and the dynamics, and can be identified with the indirect interaction of Solon et al..   Eq.  \eqref{YBGsette} in the limit $\tau \to 0$ reproduces the expression of the hydrostatic equilibrium of a standard molecular fluid in a local density approximation.
There is an interesting
analogy with granular gases subject to homogeneous shaking where the kinetic temperature can be a function of position
and regions of higher density  correspond to lower temperatures because the higher rate of  inelastic collisions
 causes a higher energy dissipation rate  \cite{goldhirsch1993clustering,marini2007}.

 \subsection{Effective pair potential}
Equation  \eqref{p2bgy} determines the pair correlation
function of the model: $\rho^{(2)}(\rr_1,\rr_2)=\bar \rho^2 g(|\rr_1-\rr_2|)$ , but it
suffers of the usual problem of  statistical mechanics of liquids since it requires the knowledge 
of higher order correlations even at linear order in the expansion in the activity parameter.
One could proceed further by assigning a  prescription to determine  the three particle correlation function $g_3$ in terms of $g(r)$  as done in the literature, but we shall not follow this approach and only consider
the low density limit  where it is possible to use it to  derive a simple expression for $g(r)$ 
 and define  an effective interaction. In fact,
the pair distribution function for a two particle system
 obtained from \eqref{probabilitydistr} reads:
   \be
  g(r)=   \exp \left(- \frac{ w(r)   +\frac{\tau}{ \gamma} [ (w'(r)]^2 
   - T_s\ln [(1+ 2 \frac{\tau}{ \gamma}  w''(r)  )(1+2\frac{\tau}{ \gamma} \frac{w'(r)}{r})^2]  } {T_s}   \right) \ee
   where the apostrophes mean derivative with respect to the separation $r$.
   The effective pair potential is defined as
$  \phi(r)=-T_s\ln g(r)$.  
  Let us remark that  this method of introducing the effective potential does not account for the three body terms
  which instead would emerge from the solution of the BGY-like equation. The lack of such contributions affects the calculation of the virial terms  beyond the second   in the pressure equation of state discussed in section \ref{vdW}.

\section{Mean field pseudo-Helmholtz functional}
\label{MFT}
As we have seen above, the BGY hierarchy is instructive, but unpractical because it requires a truncation scheme (for instance the Kirkwood superposition
approximation  at the level of two-particle distribution function  to eliminate the three body distribution) and 
 any progress can be obtained only numerically. The approximations involved  are
 often difficult to assess,  and  for the sake of simplicity  in the present paper we limit ourselves to a simpler mean field approach.
 A method often employed in equilibrium statistical mechanics to construct a mean field theory is the
	 so called variational method based on the Gibbs-Bogoliubov inequality \cite{hansen1990theory}. At equilibrium 
 one can prove that there exists a Helmholtz free energy functional, $\Fu^{eq}[f]$ of the probability density distribution, $f$, in configuration
 space such that it attains its minimum value, when the generic distribution $f$, selected among those which are normalized and non negative, corresponds to the equilibrium distribution function $f^{eq}$.
%
Such a method can be generalized to the present non equilibrium case to develop a mean field theory.
We start from the probability distribution   \eqref{probabilitydistr} and introduce the "effective Helmholtz free energy" functional as
\be
 \Fu [P_N^{trial}]\equiv \mathrm{Tr} P_N^{trial}(\Ha+T_s\ln P_N^{trial})
 \label{functional1}
\ee 
where 
$\mathrm{Tr} \equiv \iint d\rr_1 ,\dots,d\rr_N$
and $P_N^{trial}$ is an arbitrary $N$-particle normalized probability distribution, thus integrable and non negative.
Define 
\be
Z_N=\mathrm{Tr} \exp \left[- \frac{\Ha(\rr_1,\dots,\rr_N)}{T_s} \right] .
\ee
The stationary probability distribution which minimizes $\Fu$ is: 
\be
f^0_N = \exp \left[ -\frac{\Ha(x_1,\dots,x_N)}{T_s} \right] \, Z_N^{-1}
\label{fequilibrium}
\ee
and  is the analogue of the equilibrium density distribution.
 In fact, the 
effective free energy:
\be
\Fu_0 [f^0_N]\equiv-T_s\ln Z_N[f^0_N]
\label{fconnection}
\ee
is a lower bound for all distributions such that
$
\Fu [P^{trial}_N]>\Fu_0 [f^0_N]
\label{minimumfunctional}
$
for  $P^{trial}_N\neq f^0_N$
(see ref. \cite{hansen1990theory}).

In other words, the functional evaluated with any approximate distribution has a "free energy" higher than the one corresponding
to the exact distribution.
 Since $\Fu$ is minimal at the steady state it is then reasonable to assume that it can play the role
of an Helmholtz free energy in an equilibrium system. Using such an analogy we shall construct an explicit (but approximate) mean field expression
of $\Fu$ in terms of the one body distribution function $P_1$ or equivalently $\rho^{(1)}$
Let us remark that our present results do not allow to establish the central achievement of density functional theory (DFT),
	 namely the fact that the $N$-particle distribution function $f_N^0$  is a unique functional of the one body density distribution \cite{evans1979nature}. 
  This in turn means that for any fluid exposed to
an arbitrary external potential $u(\rr)$, the intrinsic free energy is a unique functional of
equilibrium single-particle density .

Nevertheless, it is possible to prove
another remarkable property    of  $\Fu$, an H-theorem,  stating  that the time derivative of $\Fu$ evaluated 
on a time dependent solution of the FPE is non positive, meaning that  the functional always decreases 
during the dynamics and tends to a unique solution. The proof of this property  is a simple application of a 
general result concerning the Fokker-Planck equation and can be found in the book by Risken \cite{risken} and in appendix \ref{Htheorem} we provide
the necessary elements to obtain it.

In order to construct 
the mean field approximation one assumes the following factorization of the $N$-particles probability distribution,
$
P^{trial}_N(\rr_1,\dots,\rr_N)= \Pi_{k=1}^N P_1(\rr_k)$,
and minimizes the functional 
\be
 \Fu [P^{trial}_N]\equiv Tr P^{trial}_N(\Ha+T_s\ln P^{trial}_N)
\ee 
since the minimum of such a functional gives the closest value to the true functional $\Fu [f_N]$ within those
having the product form as above:
\be
 \Fu [P^{trial}_N]\equiv N T_s  \int d\rr P_1(\rr) \ln P_1(\rr)+Tr P^{trial}_N \Ha \,
\ee 
having used the fact that all $P_1$ are normalized to 1.
Switching to the $\rho^{(1)}$ distribution we find:
\be
 \Fu [\rho^{(1)}]= T_s\int d\rr \rho^{(1)}(\rr)  (\ln \rho^{(1)}(\rr)-1)+\sum_1^N \langle\Ha_n\rangle
 \label{mffunctional}
 \ee 
where we considered the following decomposition of $\Ha=\sum_1^N\Ha_n$
into one body, two body, three body up to $N$-body interactions and their averages
$
\langle \Ha_n\rangle =\int\dots\int  d\rr_1\dots d\rr_n   \rho^{(1)} (\rr_1)\dots \rho^{(1)} (\rr_n) \Ha_n(\rr,\dots,\rr_n) .
$
The minimum of the functional $\Fu$ is obtained by differentiating w.r.t. to the $  \rho^{(1)} (\rr) $.
 \subsection{Application to Planar and Spherical interfaces of non interacting systems}
As an example we consider the "Helmholtz functional" for non interacting particles near a flat wall, for which only $\Ha_1$ contributes:
\bea
 \Fu [\rho^{(1)} ] \propto \int dx \rho^{(1)}(x)  \Bigl(T_s(\ln \frac{\rho^{(1)}(x)}{\rho_0}-1)-
 T_s \ln[1+ \frac{\tau}{\gamma}
  u_{xx}(x)   ] +  u(x)+ \frac{\tau}{2\gamma}
 (  u_{x}(x))^2  \Bigl)
\label{onedimfunctional}
 \eea 
 minimizing w.r.t. $\rho^{(1)}$ we obtain the same result as  \eqref{walldistribution}  which is exact.
 Similarly, the extension to three dimensional spherical walls is 
 \bea
 \Fu [\rho^{(1)}]= \int d\rr \rho^{(1)}(r)  \Bigl(T_s(\ln \frac{\rho^{(1)}(r)}{\rho_0}-1)-
 T_s \ln[(1+ \frac{\tau}{\gamma}
   u''(r)) (1+ \frac{\tau}{\gamma}
   \frac{u'(r)}{r})^2    ] +  u(r)+ \frac{\tau}{2\gamma}
 ( u'(r))^2  \Bigl)\nonumber\\
 \label{sphericalfunctional}
 \eea  
 Notice the difference of order $\tau/\gamma$, namely  a factor $2$, between the $g(r)$ found by considering two active particles and the single-particle density profile  when a spherical wall particle is pinned at the origin. Such a non-equivalence  is fingerprint
 of the off-equilibrium nature of the system and disappears as $\tau\to 0$.
 \subsubsection{Mobility and pressure tensor in spherical symmetry}

 One can derive an interesting formula starting from \eqref{sphericalfunctional}. We first extremize the functional with respect to $\rho^{(1)}(r)$ 
 (i.e. $\delta \Fu/\delta \rho^{(1)}(r)=0$ ) to obtain the profile,
 hence differentiate the result with respect to $r$ and multiply it by $\rho^{(1)}$ and arrive at the following formula:
 \be
 T_s    \frac{d}{dr}    \left(   \frac{   \rho^{(1)}(r) }     { 1+\frac{\tau}{\gamma} u_{ext}''(r)  }    \right)+ 2 \frac{T_s}{r}  \left(   \frac{   \rho^{(1)}(r) }     { 1+\frac{\tau}{\gamma} u_{ext}''(r)  }    -  \frac{   \rho^{(1)}(r) }     { 1+\frac{\tau}{\gamma} u_{ext}'(r) /r}    \right)=-\rho^{(1)}(r) u_{ext}'(r). 
 \label{hydropress}
  \ee
 On the other hand in a system of spherical symmetry the pressure tensor, $p_{\alpha\beta}(r)$,  at point $\rr$ must be of the form:
$p_{\alpha\beta}(r)=   \hat r_\alpha \hat r_\beta p_N(r)+  (\delta_{\alpha\beta}- \hat r_\alpha \hat r_\beta )p_T(r)$, where 
we have separated the
  normal (N) and tangential (T) components and $\hat {\bf r}$ denotes unit vector in the radial direction.
  In the absence of currents 
 the mechanical balance  condition dictates
 $$
  \frac{d}{dr} p_N(r)   +\frac{2}{r}(p_N(r)-p_T(r))= -\rho^{(1)}(r) u_{ext}'(r). 
  $$  
By comparing with  eq. \eqref{hydropress} we may identify the two components 
 $p_N(r)=  T_s  \rho(r) /(      1+\frac{\tau}{\gamma} u_{ext}''(r)  ) $ and
 $p_T(r)= T_s  \rho^{(1)}(r)/( 1+\frac{\tau}{\gamma} u_{ext}'(r)/r  ) $  \cite{rowlinson2013molecular}.
   It would be interesting to extend the above analysis to the case of interacting particles along the lines
  sketched in section \ref{BGY} , but this task is left for a future paper.
  Eq. \eqref{hydropress}
 provides an interesting generalization of the planar formula \eqref{YBG1t} to a spherical interface and shows that unlike 
  equilibrium fluids the normal and  tangential components of the pressure tensor of a gas of {\it non interacting} active particles
	  are not equal in the proximity of a spherical wall, but they tend to the common value $T_s \rho^{(1)}$ when $\tau\to 0$ .
	  Notice that $p_T(r)$  exceeds $p_N(r)$ for a repulsive wall-potential  since
$u_{ext}'(r)<0$. By pursuing the analogy with equilibrium statistical mechanics one could identify the integral of the second term in 
 eq. \eqref{hydropress} with a  mechanical surface tension, which by our argument would turn up to be negative in agreement with
 Bialk\'e et al. \cite{bialke2014negative}. A question arises: what is the nature of $p_N$ and $p_T$?  
  The two quantities are two components of the swim pressure tensor and are different from their common bulk value $p$,
  because near a a spherical obstacle the motilities of the particles along the normal and in the tangential plane to the surface
  are also different.  
As far as  the mobility tensor is concerned in  spherical geometry we can decompose it as 
 \be
\frac{1}{\gamma} \Gamma^{-1}_{\alpha\beta}(r)= \frac{1}{\gamma} \left[ \frac{1}{\Gamma_N(r)}  \hat r_\alpha \hat r_\beta   +\frac{1}{\Gamma_T(r)}(\delta_{\alpha\beta}- \hat r_\alpha \hat r_\beta ) \right]
 \ee
 with $ \Gamma_N(r)  =(      1+\frac{\tau}{\gamma} u_{ext}''(r)  ) $  and  $ \Gamma_T(r)  =(      1+\frac{\tau}{\gamma} u_{ext}'(r)/r  )$  
	 which shows why the tangential motion is characterised by a higher mobility than the normal motion near a curved surface as reported by many authors on the basis of simulation results and phenomenological arguments \cite{elgeti2013wall}.
 Particles are free to slide along the directions tangential to the surface and this explains why $p_T>p_N$.

\begin{figure}[h]
\centering
  \includegraphics[height=10cm]{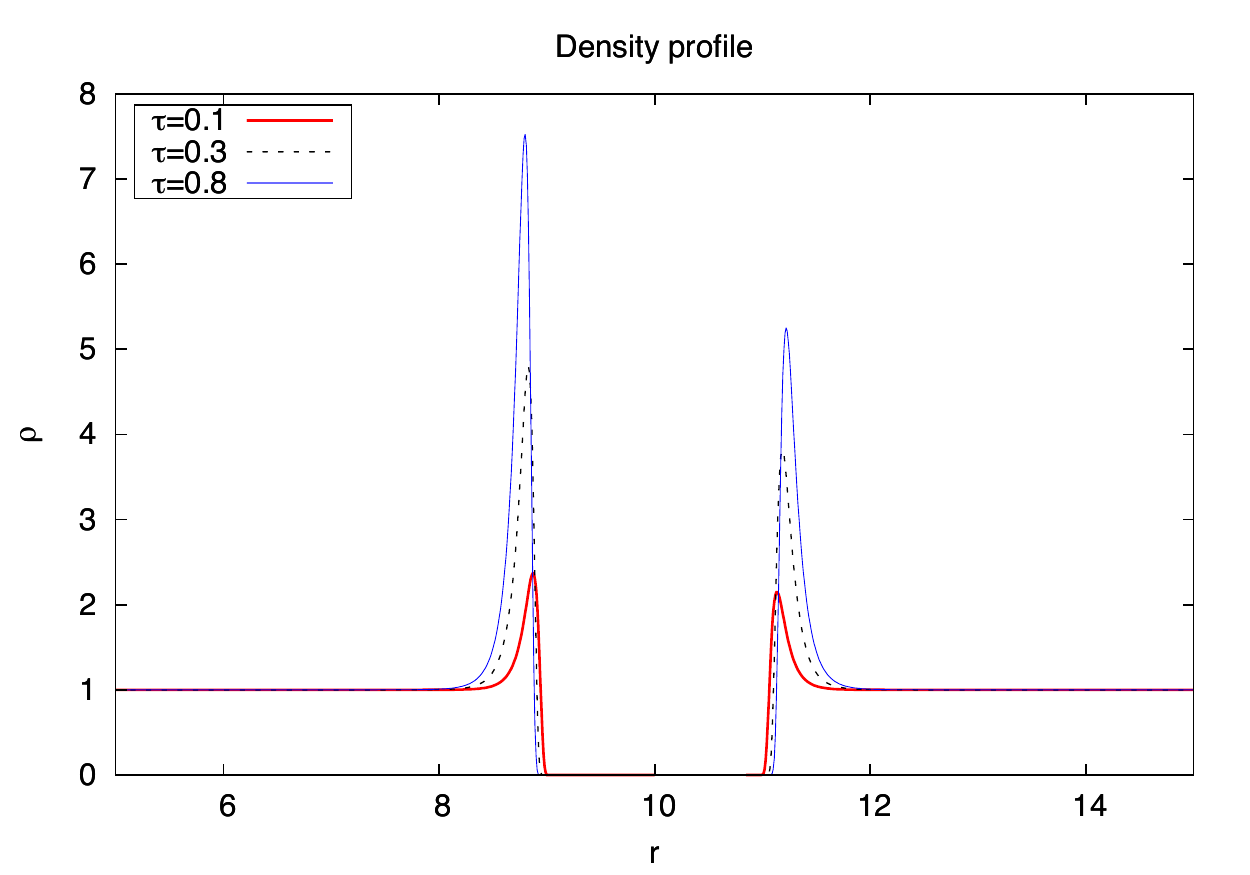}
  \caption{Comparison between the outer and the inner local density  profiles for three choices of $\tau/\gamma$  in the presence of a repulsive spherical soft wall at $R_0=10$.  Particles are non interacting.}
  \label{figconcaveprofiles}
\end{figure}

\begin{figure}[h]
\centering
  \includegraphics[height=10cm]{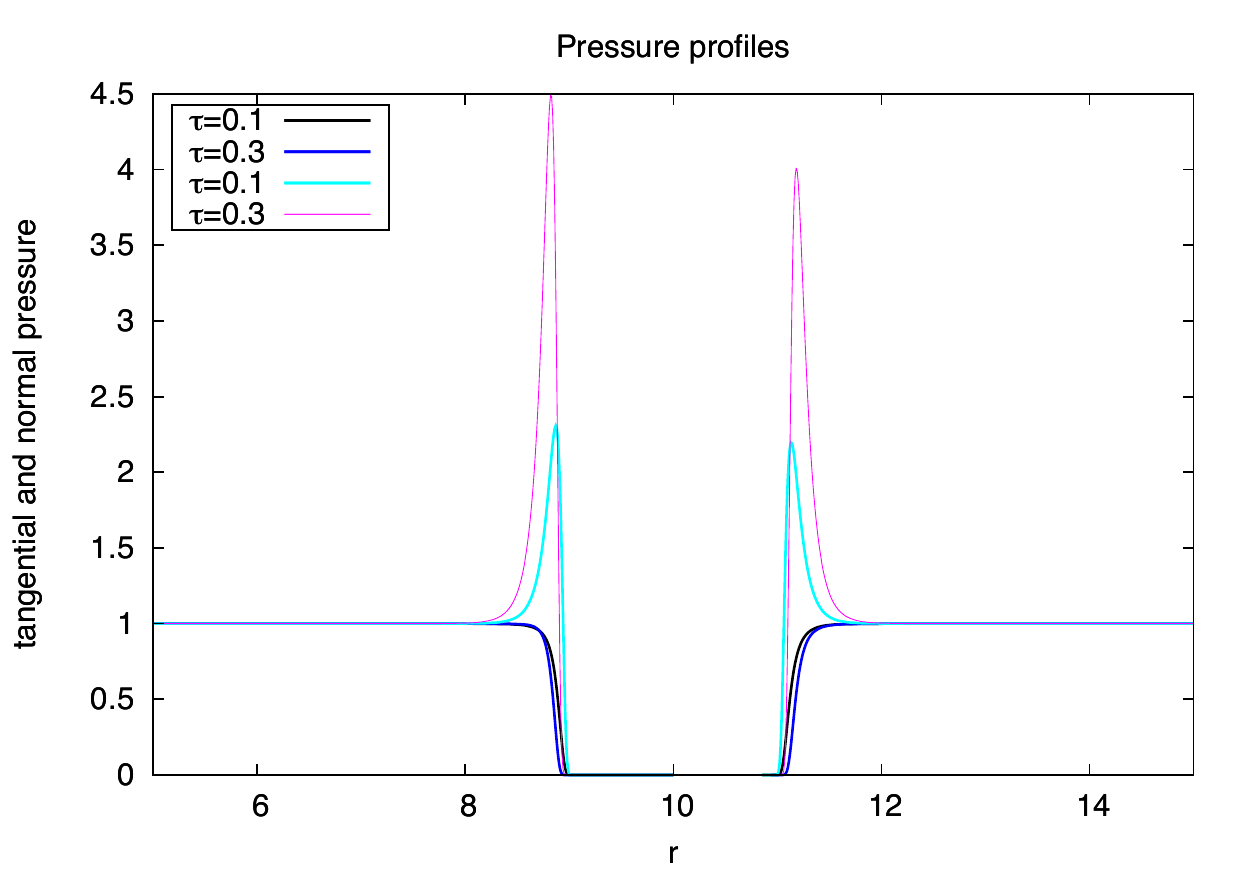}
  \caption{Comparison between the outer and the inner local pressures  profiles for two choices of $\tau/\gamma$, 0.1 and 0.3  in the presence of a repulsive spherical soft wall at $R_0=10$.  The tangential component displays a peak 
  at a distance $\approx \sigma$ from the wall, whereas 
  the normal component decreases monotonically as the wall is approached. Notice that the tangential components are  always larger than the corresponding normal components.}
  \label{figconcavepressures}
\end{figure}

 Recently, Smallenburg and Lowen   \cite{smallenburg2015swim} have studied numerically  non interacting active spheres in spherical geometry
 and found that for finite curvature radii active particles in contact with the inside of the boundary tend to spend  larger time than those 
 at the outside and differences  between the inner and the outer density profile as a function of the normal
 distance to the wall.
To see how  the density profile around a spherical repulsive wall reduces to the profile near a planar wall
   we use the explicit form of the solutions and introduce the normal distance,  $z=r-R_0$, to the spherical wall defined by the potential $u_0 \sigma^n/(r-R_0)^n$  and write for $z>R_0$:
      \bea
&&
  \rho^{(1)}(z)=  \rho_0 \exp \left[-  \frac{ 1}{T_s} \left(u_0 \left( \frac{\sigma}{z} \right)^{n}   +\frac{\tau}{ 2 \gamma}  \frac{u_0^2}{\sigma^2}  n^2 \left( \frac{\sigma}{z}\right)^{2n+2} \right)  \right]\nonumber\\
 &&
 \left(  1+  \frac{\tau}{ \gamma} \frac{u_0}{\sigma^2} n(n+1) \left( \frac{\sigma}{z} \right)^{n+2}   \right)  \left(  1- \frac{\tau}{ \gamma} \frac{ u_0}{\sigma} \frac{ 1}{R_0+z} n 
\left( \frac{\sigma}{z} \right)^{n+1}   \right)^2  . 
  \eea
   In the limit of $z/R_0\to 0$ the spherical profile must reduce to the planar profile, and indeed this is the situation as one can see  by expanding the formula above in powers of $z/R_0$ . For particles contained in a spherical cavity ($z<R_0$) the potential reads $u_0 \sigma^n/(R_0-r)^n$
 and one must replace the last factor by  $\left(  1+ \frac{\tau}{ \gamma}  \frac{ u_0}{\sigma}  \frac{ 1}{R_0-z}  n (\frac{\sigma}{z})^{n+1}   \right)^2 $.
The average local density on the concave side of a sphere is larger than the corresponding density on the convex side
as found numerically by Mallory et al.   \cite{mallory2014curvature}. This is illustrated in Fig. \ref{figconcaveprofiles} and
 Fig. \ref{figconcavepressures}.


\subsection{Can we write a density functional for interacting active particles?}
  As discussed above, since in the case of non interacting systems it is straightforward to construct an appropriate "Helmholtz" free energy functional such that it yields  the same equation as the BGY method, one would like 
 to  determine the corresponding functional also in the case of interacting particles even within the simplest mean-field 
 approximation.  To this purpose we should construct a mean field functional $\Fu$ whose functional derivative with respect to $\rho^{(1)}$ gives a non equilibrium chemical potential $\mu(\rho^{(1)})$ whose gradient must vanish in the steady state
 giving  a condition identical to the BGY equation \eqref{rho1bgy}.  We have found that such a program can be carried out
 in the one dimensional case, in fact the following construction has the required properties
  \bea
&&
 \Fu [\rho^{(1)} ] = T_s \int dx  \rho^{(1)}(x)  (\ln \frac{\rho^{(1)}(x)}{\rho_0}-1)-
 T_s  \frac{\tau}{\gamma}\int dx
  u_{xx}(x) \rho^{(1)}(x)      -\frac{1}{2}T_s \frac{\tau}{\gamma}\iint
  dx dx' \, w_{xx}(x,x')\rho^{(1)}(x) \rho^{(1)}(x') \nonumber \\
  &&
  +  \int dx\, u(x) \rho^{(1)}(x) 
  +\frac{1}{2} \iint
  dx dx' \,w(x,x')\rho^{(1)}(x) \rho^{(1)}(x')
   \, .
  \label{onedimfunctional2}
 \eea 
Unfortunately, we are not able to write the corresponding functional for dimensions higher than one, the difficulty being the
presence of the off-diagonal elements of the friction matrix  feauturing in eq. \eqref{rho1bgy} which render 
the equation non integrable. In other words, eq. \eqref{rho1bgy}  is only a mechanical balance condition and involves
transport coefficients, such as $\Gamma_{ij}$, 
thus revealing the true off-equilibrium nature  of the active system, while in passive systems the transport coefficients
never appear when thermodynamic equilibrium holds.

\section{ Van der Waals  bulk equation of state for active matter}
\label{vdW}
 
 As we have just seen the construction of a free energy functional capable of describing the inhomogeneous
 properties of active particles remains an open problem, nevertheless one can focus on the less ambitious task of determining the
 pseudo-free energy of a bulk system
 characterized by an effective pair interaction $\phi(r)$, defined in Section \ref{BGY}.
 Notice that in the homogeneous case  the friction matrix reduces to a scalar
 quantity due to the higher symmetry.
 As a prerequisite for a successful mean-field approximation 
 $\phi(r)$  must be splitted into a short range repulsive part, $\phi_{rep}$,  and a weaker longer range  attractive contribution,  $\phi_{at}$,
  and only the latter can be treated in a mean field fashion, since the effect of repulsion is a highly correlated phenomenon
  and not perturbative.
 Explicit expressions for the splitted potentials are given by \eqref{phirep} and \eqref{phiat}.
To capture the effect of repulsive forces  a simple modification of the ideal gas entropic term, already
 introduced by van der Waals to account for the reduction of configurational entropy due to the finite volume
 of the particles, is sufficient.
Thermodynamic perturbation theory would represent the natural choice to determine the total free energy.  It assumes
 a reference hard-sphere system, whose Hamiltonian only depends on $\phi_{rep}$, as the unperturbed system. In particular the reference
system characterized by a temperature dependent 
 diameter, $d(T)$,  allows to determine the free energy excess associated with the perturbing potential $\phi_{at}$ and
 and equation of state of the model.
However, since perturbation theory still requires a considerable amount of computer calculations, here we make  the simplest ansatz  and write the following free energy functional
\bea
\Fu_{vdW} =T_s \int dr \rho^{(1)}  (\rr) \left[\ln \left(
\frac{\rho^{(1)} (\rr)}  {   1- \frac{2\pi}{3}\rho^{(1) }(\rr)  d^3(T_s)}\right)-1 \right] 
+ \frac{1}{2}\int d\rr \int_{|\rr'-\rr|>d} d\rr' \phi_{at}(\rr-\rr') \rho^{(1) }(\rr) \rho^{(1) }(\rr') 
\label{fuvdw}
.  \eea
The first term is just the entropy of a fluid with the excluded volume correction, the second term stems from the activity and may lead to condensation phenomena for large values of $\tau$. In relation to the naive mean-field functional \eqref{onedimfunctional2}, the van der Waals
entropic term already contains the repulsive part of the direct interaction, whereas the attractive term which vanishes in the 
$\tau\to 0$ limit takes into account the activity.
In principle this 
 form of  free energy can be constructed by employing the Mayer cluster  expansion to evaluate $Z_N$, associated with the
 effective Hamiltonian $\Ha$,
in the approximation of neglecting $m$-body interactions with $m>2$ there contained, that is using the effective potential $\phi(r)$.
With these ingredients one can define the active component  of the van der Waals attractive parameter:
\be
a^{(A)}=-(1/2)\int_{r>d(T_s)} d^3 r \phi_{at}(r)
\label{a_active}
\ee
and 
the covolume 
$$b=2\pi \frac{d^3(T_s)}{3}$$ and estimate  $\Fu_{vdW}$. Since the procedure adopted is completely analogous to the
one employed in equilibrium fluids 
one can immediately derive a pressure equation by differentating $\Fu_{vdW}$ with respect to volume:
$$
p_{vdW} =-\frac{\partial  {\Fu_{vdW}}}{\partial V} =T_s\frac{\rho}{1-b\rho}-a^{(A)} \rho^2
$$
which at low density reduces to the ideal bulk swim pressure $T_s \rho$.
Going further, one can add a
square gradient contribution to the free energy density, that  simplifies the non local functional \eqref{fuvdw}
while allowing to describe inhomogeneous systems. This is achieved by introducing a term 
$
m|\nabla \rho|^2
$,
where $m=-(1/12)\int_{r>d(T_s)} d^3 r r^2 \phi_{at}(r)$ \cite{evans1979nature}.
   We choose, now, the bare potential of the form
  $w(r)=w_0(\frac{\sigma}{r})^\alpha$, where $w_0$ is the strength of the bare potential, $\sigma$ is a nominal diameter
  and define $\tilde r=r/\sigma$, the non dimensional temperature  $T^*=\frac{T_s}{w_0}$  and the parameter
  $K_p=\frac{w_0\tau}{\gamma\sigma^2}= \frac  {(1+D_t/D_a)}{T^* } Pe^2$, where the last expression displays the
  dependence on  the P\'eclet number.
We, now, set :
\be
\frac{\phi_{rep}(\tilde r)}  { T_s}= \frac{1}{T^*}   
\left( \frac{1}{\tilde r^{\alpha}}+ K_p 
\frac{\alpha^2 }{\tilde r^{2 \alpha+2}}
\right)
\label{phirep}
\ee
and 
\be
\frac{\phi_{at}(\tilde r)}{T_s}=  -\ln \left( \left[ 1+2 K_p\frac{\alpha(\alpha+1)}
{\tilde r^{\alpha+2}} \right] \left[1- 2 K_p  \frac{\alpha}{\tilde r^{\alpha+1 }}
\right]^2 \right)  
\label{phiat}
 \ee  
for $\tilde r>1$.
The effective hard-sphere diameter is given by the Barker-Henderson formula:
\be
\frac{d(T_s)}{\sigma}=\int_0^\infty d\tilde r (1-e^{-\phi_{rep}(\tilde r)/T_s})
\ee
and is density independent, but depends on the parameter $K_p$ and temperature.
As $K_p\to 0$ (i.e. at small P\'eclet number, i.e. small persistence time ) the system approaches the behavior of an assembly of passive 
soft spheres at temperature $T^*$,
 and inverse power law potentials analytic expressions for $d(T_s)$ exist,
 whereas as $K_p\to\infty$ (large P\'eclet number) the  activity becomes increasingly relevant
 and the effective radius increases with $K_p$.

 Let us remark
that with the  definition \eqref{phiat} of  attractive potential the coefficient $a^{(A)}$ of formula \eqref{a_active} has a strong dependence on the temperature $T_s$, at variance with standard passive fluids where the dependence occurs through $d(T_s)$ only. This makes
the condensation transition basically athermal and driven by the activity parameter.
This feature renders the case of active particles possessing also a direct attractive potential contribution of the 
Lennard-Jones 6-12 type particularly interesting, that is obtained by modifying the bare pair potential as $w_{\lambda}(r)=w_0[(\frac{\sigma}{r})^\alpha
-\lambda(\frac{\sigma}{r})^{\alpha/2}] $. 
In fact, one has to consider a passive contribution similar to \eqref{a_active} to the vdW coefficient, say $a^{(P)}$ whose dependence on temperature
is rather weak as compared to $a^{(A)}$.  Thus one can expect that at low temperature and small $Pe$ the condensation is driven by
standard attraction, whereas at high $T_s$ but large $Pe$ the transition is determined by the effective attractive 
determined by the activity \cite{redner2013reentrant}.

We turn, now, to consider the  virial  series of the pressure:
$$
\frac{p }{T_s \rho}=(1+B(\tau) 
\rho+ C \rho^2+\dots)
$$
and by comparing with the van der Waals equation 
we obtain second virial coefficient $B(\tau)= (\frac{2\pi}{3}\rho d^3  -a^{(A)}(\tau)/T_s)$. 
Using the effective pair potential  we can obtain $B$ as the integral (with $\lambda=0$)
\bea
&&
B(T^*,K_p)=2\pi\sigma^3  \int_0^\infty d\tilde r \tilde r^2   \Bigl \{1-\exp\Bigl[-  \frac{1}{T^*}   
\Bigl( \frac{1}{r^{*\alpha}}+K_p
\frac{\alpha^2}{\tilde r^{2 \alpha+2}}\Bigl) \Bigl]
\nonumber\\
 &&
 \left[ 1+2 K_p
\frac{\alpha(\alpha+1)}{\tilde r^{\alpha+2}} \right] 
\left[1-2 K_p  \frac{\alpha}{\tilde r^{\alpha+1}} \right]^2 \Bigr\}.
\label{repulsive}
\eea
Formula \eqref{repulsive} gives the second virial coefficient for active spheres with purely repulsive potential
a straightforward extension of it yields the values with finite values of $\lambda$. 

\begin{figure}[h]
\centering
  \includegraphics[height=10cm]{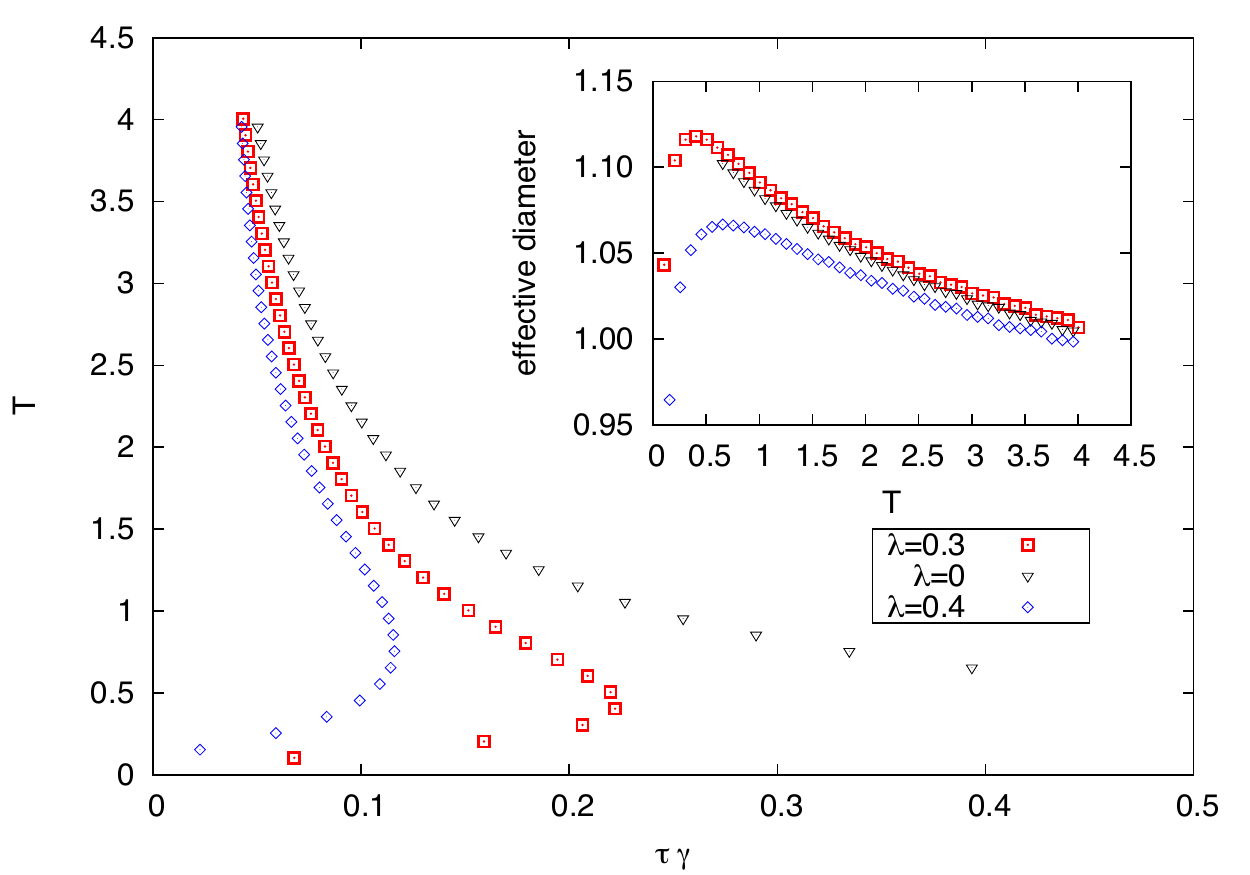}
  \caption{ Plane $\tau,T_s$: the three curves in the main panel correspond to three different values (from right to left) of the attractive
  parameter $\lambda=0,0.2,0.4$ and $\alpha=12$, respectively, and represent the loci where the second virial coefficient,  $B(\tau,T_s)$, 
  vanishes. In the purely repulsive case (triangles) the behavior is monotonic, whereas when attraction is present
  the lines bend  in the low temperature region, where B changes sign under the effect of the direct attractive force.
  In the inset we plot the effective Barker-Henderson diameter as a function of the effective temperature measured along the Boyle line in the three cases.
 }
  \label{boyle}
\end{figure}

The {\it Boyle activity parameter}, $\tau^*$  of the model, corresponds to the values where $B=0$.
For systems without attraction the curve $T_s(\tau^*)$ where $B(T_s,\tau^*)=0$ is monotonic and decreases when $\tau^*$ increases
and the resulting
 phase diagram in a plane $\tau,T_s$ is shown in Fig \ref{boyle}: the region at the right of each line corresponds to $B<0$ 
 for that given value of $\lambda$.
The same calculation is repeated  for an  active system {\it with attraction} and we found that the locus where $B(T_s,\tau^*)=0$ 
initially decreases as a function of $\tau$, but then it
bends
as displayed in Fig.   \ref{boyle} and shows at sufficiently low temperatures $T_s$ shows the presence of a region close to the origin
where $B<0$.
Figure  \ref{boyle}  displays a reentrant behavior of the Boyle's line for $\alpha=12$ when $\lambda>0$: at low temperatures $T^*$ 
the region $B<0$ under the effect of the direct attractive force  extends at small values of $\tau$, while the bending is totally
absent in the case $\lambda=0$.   The Barker-Henderson diameter  diplays the same non monotonic trend along the Boyle line
showing the correlation between the structural properties and the thermal properties of the system.

\section{Conclusions}
 \label{conclusions}
In this paper we have investigated using a microscopic approach the steady state properties of a system of 
active particles and
determined the stationary $N$-particle probability distribution function   by requiring the vanishing of all currents.
We  used such a distribution to construct the BGY hierarchy for the reduce $n$-particle (with $n<N$) distribution functions and found that
it  contains terms which do not have a counterpart in   systems of passive particles.
The presence of these terms explains several experimental or numerical findings such the modified barometric law,
the effective attraction between nominally repulsive objects, negative second  virial coefficient, condensation.
We defined an  effective potential of attractive character due to the reduced mobility experienced by  the active particles and determined by the combined effect of the
excluded volume and the persistence of their motion. In an inhomogeneous environment the mobility  turns out to be anisotropic and described by a tensor, whose
expression we have obtained in the case of a spherical surface by separating its normal and tangential components. The same type of analysis allows to determine
the mechanical pressure in systems of non interacting active particles via an hydrostatic balance, but the full extension to the interacting case remains
an open problem.
In order to obtain a bulk equation of state connecting pressure to density, swim temperature and activity parameter we have
explored an alternative approach and constructed
a mean field theory introducing   a  (pseudo) Helmholtz functional  and determined the stationary distribution by a variational principle.
In spite of the great similarities with equilibrium systems, profound differences remain.  Already at the level of the first BGY 
equation one sees that it is not possible to disentangle the interaction terms stemming from the external potential
from those from the particle-particle interactions. Such a feature in our opinion seems to obstruct the way to 
establish a full density  functional theory. In fact, it does not seem possible to prove the so called v-representability
property, i.e. the unique correspondence between the one particle density profile and a given external potential, which is at the basis
of DFT.  On the hand, if one constructs the functional using  the two body effective potential $\phi(r)$ 
the DFT approach sounds promising to tackle the properties of active particles under inhomogeneous condition.
Finally we note that the homogeneous free energy that we have constructed seems to reproduce the behaviour of the active fluid, helps in locating its instabilities and the onset of phase separation. In this context it would be very interesting to compare our approximate theory with numerical simulations of interacting colored-noise driven particles.
In the future  we plan to apply the  pseudo-Helmholtz functional to the analysis of strongly in-homogeneous systems of active particles
and to investigate the MIPS in these situations.

\section*{Acknowledgments}
We thank A. Crisanti,  L. Cerino, S. Melchionna, A. Puglisi, A. Vulpiani and R. Di Leonardo for illuminating discussions.
C. Maggi acknowledges support from the European Research Council under the European Union's Seventh Framework programme
(FP7/2007-2013)/ERC Grant agreement  No. 307940.

\appendix

\section{Multidimensional unified colored noise approximation}
\label{appendixA}

In the present appendix, following the method put forward by Cao and coworkers \cite{cao1993effects,ke1999phase}, we
introduce  an auxiliary stochastic process, $w_i$, defined by:
$$
w_i=\frac{F_i}{\gamma}+v_i .
$$
By taking the 
 derivative of $w_i$ with respect to $t$ we obtain
\be
\dot  w_i = \frac{1}{\gamma} \sum_k \frac{\partial F_i}{\partial x_k} \dot x_k+\dot v_i .
\label{e55}
\ee
after substituting into eq. \eqref{e55}  relations   \eqref{effective_langevin}and  \eqref{exp_correlation}   to eliminate $\dot v_i$ and $\dot x_k$ we arrive at the evolution equation for $w_i$:
\be
\dot w_i = \frac{1}{\gamma} \sum_k \frac{\partial F_i}{\partial x_k} (w_k+ D_t^{1/2} \xi^t_k) 
-\frac{1}{\tau}  \left[ w_i-\frac{F_i}{\gamma} \right]       +\frac{D_a^{1/2}}{\tau}\eta_i(t) .
\ee
Now, we assume the unified colored noise approximation (UCNA) \cite{hanggi1995colored} consisting in dropping the "acceleration term", $\dot w_i$,  featuring in the l.h.s. of this equation, so that we can write the following system of algebraic linear equations
for the quantities $w_i$:
\be
 \left[ \delta_{ik}- \frac{\tau}{\gamma} \sum_k \frac{\partial F_i}{\partial x_k} \right] w_k=\frac{\tau}{\gamma}   \sum_k   \frac{\partial F_i}{\partial x_k} D_t^{1/2} \xi^t_k
  +\frac{F_i}{\gamma}      +D_a^{1/2} \eta_i(t) .
\ee
With the help of the matrix $\Gamma_{ik}$ (defined by eq.\eqref{gammamatrix})
we go back to the equation for $x_i $   eq.\eqref{effective_langevin} , rewritten as $w_i=\dot x_i- D_t^{1/2}\xi_i^t$,  to find
the following Langevin equation:
\be
\dot x_i \simeq \sum_k \Gamma^{-1}_{ik}\frac{1}{\gamma}  F_k+
D_a^{1/2}  \Gamma^{-1}_{ik}\eta_k(t)  +D_t^{1/2}       \xi^t_i(t) 
+D_t^{1/2} \frac{\tau}{\gamma}  \sum_k \Gamma^{-1}_{ik}   \sum_l   \frac{\partial F_k}{\partial x_l}  \xi^t_l
\ee
that we interpret using the Stratonovich convention \cite{gardiner1985stochastic}.
We finally observe that the last two terms can be gathered together and after some manipulations reported hereafter
 \bea
 &&
 \sum_l \Bigl[ \delta_{il} + \sum_k \Gamma^{-1}_{ik}     \frac{\tau}{\gamma} \frac{\partial F_k}{\partial x_l}  \Bigl] \xi^t_l 
= \sum_l  \sum_k \Bigl[ \Gamma^{-1}_{ik} \Gamma_{kl} + \Gamma^{-1}_{ik}     \frac{\tau}{\gamma} \frac{\partial F_k}{\partial x_l}  \Bigl] \xi^t_l = \nonumber\\
&&
= \sum_l  \sum_k \Gamma^{-1}_{ik} \Bigl[\Gamma_{kl} +\frac{\tau}{\gamma} \frac{\partial F_k}{\partial x_l}  \Bigl] \xi^t_l 
= \sum_l  \sum_k \Gamma^{-1}_{ik} \delta_{kl}   \xi^t_l  =   \sum_k \Gamma^{-1}_{ik}    \xi^t_k    
\nonumber
\eea
give rise to  the  form of the stochastic equation for $x_i$  reported in eq.\eqref{stochasticxi}.


We limit ourselves to consider the stationary solutions with vanishing current 
and since we have assumed that the determinant is non vanishing the only solution of eq. \eqref{Ncurrent} corresponding to $J_l=0$   is
obtained by imposing the  conditions:
\be
\frac{1}{(D_a+D_t)\gamma} F_k P_N-P_N\sum_j \frac{\partial}{\partial x_j} \Gamma^{-1}_{jk} =\sum_j \Gamma^{-1}_{jk} \frac{\partial}{\partial x_j} P_N
\label{zerocondition}
\ee 

 Multiplying  by  $ \Gamma_{nk} $ and summing over $k$
after  some algebra  we obtain the first order differential equations for $P_N$:
\be
P_N\frac{\tau}{(D_a+D_t) \gamma} \sum_k \Gamma_{nk} F_k + P_N \sum_{jk}  \Gamma^{-1}_{jk} \left( \frac{\partial}{\partial x_j} \Gamma_{kn} \right) 
= \frac{\partial P_N}{\partial x_n}
\label{pformula}
\ee
Since the matrix elements $\Gamma_{ij}$ contain 2-nd cross derivatives $\partial^2/\partial x_i\partial x_j$ of the potential $\UU$ we have
\be
\frac{\partial}{\partial x_j} \Gamma_{kn}=\frac{\partial}{\partial x_n} \Gamma_{kj}
\ee
and we finally get
 \be
  \sum_{jk}  \Gamma^{-1}_{jk} \left( \frac{\partial}{\partial x_j} \Gamma_{kn} \right) = 
 \sum_{jk}  \Gamma^{-1}_{jk} \left( \frac{\partial}{\partial x_n} \Gamma_{kj}\right)
 = \frac{1}{\det \Gamma} \frac{\partial}{\partial x_n} \det \Gamma
 \label{Jacobi1}
 \ee
 The last equality derives from the following Jacobi's formula 
 \be
  \frac{1}{\det \Gamma} \frac{\partial}{\partial y} \det \Gamma = \mathrm{Tr} 
  \left( \Gamma^{-1} \frac{\partial}{\partial y} \Gamma \right)
  \label{jacobiformula}
  \ee
  where $y$ stands for any of the variables $x_i$.

  Explicitly by using \eqref{pformula} and \eqref{Jacobi1}  we find the  system of  differential equations \eqref{pformul2abis}  determining     the full probability distribution.  
   
   
    \section{H-theorem for the FPE}
    \label{Htheorem}
The functional $\Fu [f_N]=  Tr f_N (\Ha+T_s\ln f_N) $ discussed in Sect. \ref{MFT}  has also an interesting dynamical property. In fact,
after rewriting it under the form
$$\frac{1}{T_s}\Fu [f_N]= Tr f_N \ln \left( \frac{f_N}{P_N} \right)- \ln Z_N $$ 
 where  $f_N(\{x\},t)$ is a normalized solution of the dynamical Fokker-Planck equation  at instant $t$ and $P_N$ is the stationary distribution
 and the first term is the so called Kullback-Leibler relative entropy \cite{risken,kullback1951information}.
One  can show the following H-theorem
\be
\frac{d \Fu}{d t}\leq 0 .
\ee

The proof follows closely the method presented by Risken \cite{risken} (section 6.1 of his book), but before proceeding it is necessary to reduce the FPE described by  eqs.
\eqref{FPE} and \eqref{Ncurrent} to the canonical form:
\be
\frac{ \partial f_N}{\partial t}=\sum_{i}  \frac{\partial }{\partial x_i} \Bigl[\sum_j  \frac{\partial }{\partial x_j} D_{ij}^{(2)} f_N   -D_i^{(1)} f_N\Bigr]
\label{fNevolution}
\ee
where the diffusion matrix is
\be
D_{ij}^{(2)} =(D_a+D_t)\, \Gamma^{-2}_{ij}
\label{d2ij}
\ee
and the drift vector is
\be
D_i^{(1)}=  \frac{1}{\gamma } \Gamma^{-1}_{lk}\, F_k+(D_a+D_t) \Gamma^{-1}_{jk}\, \frac{\partial }{\partial x_j} \Gamma^{-1}_{ik}
\label{d1i}
\ee
Now using some standard analytic manipulations reported by Risken it is simple to show the following formula
\be
\frac{d \Fu}{d t}= -\int d^N x f_N  D_{ij}^{(2)} \,  \frac{\partial \ln R}{\partial x_i} \frac{\partial \ln R}{\partial x_j}  \leq 0 
\ee
where $R=f_N/P_N$. If   $D_{ij}^{(2)} $ is positive definite $\Fu$ must always decrease for $\frac{\partial \ln R}{\partial x_i}\neq 0$
towards the minimum value $-T_s \ln Z_N$. This result also implies that the  solution of the FPE is unique, and after some 
time T the distance between two solutions is vanishingly small.
\section{Detailed Balance}
\label{detailedbalance}
The detailed balance implies a stronger condition than the one represented by having a stationary distribution,
since it implies that there is no net flow of probability around any closed cycle of states. In practice if
detailed  balance holds it is not possible to have a ratchet mechanism and directed motion.
 Again we use the equations derived by Risken (Section 6.4 of his book)\cite{risken}  which represent the sufficient and necessary conditions
for detailed balance. Since the variables $\{x_i\}$ are even under time reversal 
	we have to verify the validity of the following equations \cite{van1957derivation,graham1971generalized}:
\be
	D_i^{(1)}\, P_N(\{x\})=    \frac{\partial }{\partial x_j} D_{ij}^{(2)} \,  P_N(\{x\})
\ee
Using the explicit form of \eqref{d2ij} and \eqref{d1i} it is straightforward to verify
 that the detailed balance conditions are verified since they are the same as the condition  expressing the vanishing of the 
 current components $J_i$ in the stationary state (see eq. \eqref{fNevolution}.


 \section{Approximation for the determinant}
 \label{appendixC}
 The exact evaluation of the determinant $\Gamma$ associated with the Hessian matrix 
 is beyond the authors capabilities and 
  we look for approximations in order to evaluate the effective forces.
 We consider
the associated determinant in the case of two spatial dimensions:

$$
\begin{small}
\left(\begin{array}{ccccccc}
[1+\frac{\tau}{\gamma}\sum_{j\neq 1} w_{xx}(\rr_1,\rr_j)]& \sum_{j\neq 1} \frac{\tau}{\gamma}w_{xy}(\rr_1,\rr_j)& -\frac{\tau}{\gamma}w_{xx}(\rr_1,\rr_2) &\dots&-\frac{\tau}{\gamma}w_{xy}(\rr_1,\rr_N) 
\\ \sum_{j\neq 1} \frac{\tau}{\gamma}w_{yx}(\rr_1,\rr_j)   & [1+\frac{\tau}{\gamma}\sum_{j\neq 1} w_{yy}(\rr_1,\rr_j)] & -\frac{\tau}{\gamma}w_{yx}(\rr_1,\rr_2)&\dots     &-\frac{\tau}{\gamma}w_{yy}(\rr_1,\rr_N)
\\ -\frac{\tau}{\gamma}w_{xx}(\rr_2,\rr_1)& -\frac{\tau}{\gamma}w_{yx}(\rr_2,\rr_1) & [1+\frac{\tau}{\gamma}\sum_{j\neq 2}  w_{xx}(\rr_2,\rr_j)] & \dots&-\frac{\tau}{\gamma}w_{xy}(\rr_2,\rr_N) \\
\\\dots&\dots&\dots&\dots&\dots&\\
\\ -\frac{\tau}{\gamma}w_{xy}(\rr_N,\rr_1)& -\frac{\tau}{\gamma}w_{yy}(\rr_N,\rr_1)&\dots & \dots& [1+\frac{\tau}{\gamma}\sum_{j\neq N}  w_{yy}(\rr_N,\rr_j)] 
\end{array}\right)
\end{small}
$$
and  to order $\tau/\gamma$ is 
\be
 \det \Gamma\approx1+ \frac{\tau}{\gamma} \sum_{i,j,i\neq j} [w_{xx}(\rr_i,\rr_j)+w_{yy}(\rr_i,\rr_j)  ]
\ee
It is interesting to remark that the off-diagonal elements contain only one term, while the diagonal elements and their
neighbors contain $N$ elements. Thus in the limit of $N\to\infty$ we expect that the matrix becomes effectively diagonal. 
 
\bibliography{theorycolored.bib} 

\end{document}